%% file: chen_WCL2021-0665.tex
\documentclass[journal]{IEEEtran}
\usepackage{ifpdf}
\usepackage{url}
\ifpdf
\else
\fi

\usepackage{cite}
\ifCLASSINFOpdf
  \usepackage[pdftex]{graphicx}
  \graphicspath{{../pdf/}{../jpeg/}}
  \DeclareGraphicsExtensions{.pdf,.jpeg,.png}
\else
  \usepackage[dvips]{graphicx}
  \graphicspath{{../eps/}}
  \DeclareGraphicsExtensions{.eps}
\fi
\usepackage{amsmath}
\usepackage{algpseudocode}% http://ctan.org/pkg/algorithmicx
\algtext*{EndWhile}% Remove "end while" text
\algtext*{EndIf}% Remove "end if" text
\algtext*{EndFor}
\usepackage{algorithm}
\usepackage{array}

\usepackage{color}
\ifCLASSOPTIONcompsoc
 \usepackage[caption=false,font=normalsize,labelfont=sf,textfont=sf]{subfig}
\else
 \usepackage[caption=false,font=footnotesize]{subfig}
\fi
\usepackage{stfloats}
\usepackage[hidelinks]{hyperref}
\usepackage{xcolor}
\include{macros}

\makeatletter
\def\ps@IEEEtitlepagestyle{%
\def\@oddfoot{\mycopyrightnotice}%
\def\@evenfoot{}%
}
\def\mycopyrightnotice{%
{\scriptsize 2162-2337~\copyright~2021 IEEE. Personal use is permitted, but republication/redistribution requires IEEE permission. See http://www.ieee.org/publications\_standards/publications/rights/index.html.} % Revise this line accordingly!
\gdef\mycopyrightnotice{}
}

\makeatletter
\let\old@ps@headings\ps@headings
\let\old@ps@IEEEtitlepagestyle\ps@IEEEtitlepagestyle
\def\confheader#1{
\def\@oddhead{\strut\hfill#1\hfill\strut}%
}
\makeatother
\confheader{\scriptsize \shortstack{This article has been accepted for publication in a future issue of this journal, but has not been fully edited. Content may change prior to final publication. \\Citation information: DOI 10.1109/LWC.2021.3102402, IEEE Wireless
Communications Letters.}}

\begin{document}

\title{DOA Estimation with Non-Uniform Linear Arrays: A Phase-Difference Projection Approach}

\author{Hui Chen, Tarig Ballal, and Tareq Y. Al-Naffouri
% Michael~Shell,~\IEEEmembership{Member,~IEEE,}
%         John~Doe,~\IEEEmembership{Fellow,~OSA,}
        % and~Jane~Doe,~\IEEEmembership{Life~Fellow,~IEEE}% <-this % stops a space
\thanks{The authors are with the Division of Computer, Electrical and Mathematical Science \& Engineering, King Abdullah University of Science and Technology (KAUST), Thuwal, 23955-6900, KSA. e-mail: (\{hui.chen; tarig.ahmed; tareq.alnaffouri\}@kaust.edu.sa).}}% <-this % stops a space
% \thanks{J. Doe and J. Doe are with Anonymous University.}% <-this % stops a space

% \IEEEoverridecommandlockouts
% \IEEEpubid{\makebox[\columnwidth]{978-1-5386-5541-2/18/\$31.00~\copyright2018 IEEE \hfill}
% \hspace{\columnsep}\makebox[\columnwidth]{ }}
\maketitle
% \IEEEpubidadjcol

\begin{abstract}
Phase wrapping is a major problem in direction-of-arrival (DOA) estimation using phase-difference observations. For an antenna pair with an inter-antenna spacing greater than half of the wavelength ($\lambda/2$) of the signal, phase wrapping occurs at certain DOA angles leading to phase-difference ambiguities. Existing phase unwrapping methods exploit either frequency or spatial diversity. These techniques work by imposing restrictions on the utilized frequencies or the receiver array geometry. {In addition to these restrictions, sensitivity to noise and calibration errors is another limitation of these methods.} We propose a grid-less \emph{phase-difference projection} (PDP) DOA algorithm to overcome these issues. The concept of \emph{wrapped phased-difference pattern} (WPDP) is introduced, which allows us to compute most of the parameters required for DOA estimation in an offline manner. This results in a superior computational speed in real-time compared to methods with similar DOA estimation performance. Simulation results demonstrate the excellent performance of the proposed algorithm, both in terms of accuracy and speed.

% In addition, the utilization of a pre-computed WPDP also acts as a strong prior that allows the proposed algorithm to leverage information related to the source-array geometry to improve DOA estimation accuracy.
% The sufficient and necessary condition is derived, and a simple criterion is proposed to evaluate the feasibility of implementing this algorithm. An indicator is also defined to evaluate the performance of systems with different parameters.
% The simulation shows that the proposed algorithm performs better than the benchmark algorithms for small arrays, and it is close to the CLRB.
\end{abstract}


\IEEEpeerreviewmaketitle

\section{Introduction}
\label{sec:intro}
Direction-of-arrival (DOA) estimation is an important topic for applications such as wireless sensor networks~\cite{1-wsn}, indoor positioning and tracking~\cite{3-indoor_tracking}, Radar~\cite{radar_doa}, wireless communications~\cite{2-communicaion}, and so on. 
Many DOA estimation methods have been proposed over the years~\cite{4-doa_summary,doa_book}, focusing largely on uniform linear array (ULA) configurations. 
% Many DOA estimation methods have been proposed over the years, as reviewed in~\cite{4-doa_summary,doa_book}. The literature focuses largely on the narrow-band (or single-frequency) case with uniform linear array (ULA) configurations. Extension to multi-frequency scenarios is usually straightforward.
On the other hand, the use of non-uniform linear arrays (NULAs) is also popular. NULA configurations are often utilized to extend the array aperture and consequently improve the array's DOA resolution.

{In this paper, we focus on single-source DOA estimation with NULAs. This scenario is motivated by a mmWave/THz multiple-input multiple-output (MIMO) communication context. In these systems, flexible arrays are adopted to alleviate the high computation and hardware costs (see switch-based MIMO~\cite{switches} and array-of-subarray (AOSA) structures~\cite{aosa}). When an antenna/subarray selection algorithm is applied, a NULA structure will be formed. 
In addition to the NULA structure, the analog beamforming employed in these systems can reduce multipath and provide a dominant line-of-sight (LOS) signal~\cite{sarieddeen2020overview}.
Hence, DOA estimation for a single source observed at a NULA is an important problem for mmWave/THz MIMO systems.}  

%{\color{red} Multiple-input multiple-output (MIMO), as an essential technology for mmWave/THz band communications, reduces interference and provides dominant line-of-sight (LOS) via beamforming~\cite{sarieddeen2020overview}. To alleviate the high computational and hardware cost, flexible arrays (e.g., switch-based MIMO~\cite{switches}, array-of-subarray (AOSA) structures~\cite{aosa}) can be adopted with corresponding antenna/subarray selection algorithms. In addition, the selection procedure also forms non-uniform linear arrays (NULAs) with larger apertures than those of uniform linear arrays (ULAs), which improves the DOA resolution.
%In this letter, we focus on the DOA estimation for NULAs in a single-source scenario.}

The DOA estimation problem can be formulated as an optimization of a cost function over a feasible DOA range. Usually, the process requires evaluating the cost function for the whole DOA range, searching for that function's optimum. Maximum likelihood estimation (MLE)~\cite{crlb_threshold_region} and MUSIC~\cite{13-MUSIC} are two widely used methods that exemplify this approach. A drawback of this approach is that the search process can increase the computational complexity, especially when high spatial resolution is desired.

Time-delay estimation is a fast alternative solution to DOA estimation that can produce a DOA estimate directly without applying a grid search.~\cite{timedelay_new}.
% ~\cite{6-cross_correlation,11-tarig3,timedelay_new}
The linear relationship between time delay and phase-difference makes it possible to utilize phase-difference measurements for DOA estimation. Phase-difference based DOA estimation has been reported as an effective approach for multi-carrier signals~\cite{Chen2018, pdp_eusipco2019}. Nevertheless, phase-difference based DOA estimation suffers from the occurrence of \emph{phase wrapping}
~\cite{10-tarig2,review_ref2}.
% ~\cite{10-tarig2,review_ref1,review_ref2}.

The issue of phase wrapping can be resolved by exploiting the frequency diversity available in multi-frequency signals, or by leveraging spatial diversity in single-frequency scenarios~\cite{9-tarig-frequency}. Examples of spatial-diversity phase unwrapping methods include~\cite{10-tarig2}, and {more recent off-grid approaches such as 2Q-order difference-set~\cite{2q-order} and two-step offset correction~\cite{ma2019off}}. A major drawback of these methods is that they require a specialized antenna setup. Besides, these methods tend to be sensitive to the phase noise effect.

This paper proposes a \emph{phase-difference projection} (PDP) method for DOA estimation using non-uniform linear arrays. We capitalize on a novel concept of a \emph{wrapped phased-difference pattern} (WPDP). The proposed method can be applied to an arbitrary linear array configuration of three or more sensors. {Simulation results demonstrate that the proposed method offers a good trade-off between computational complexity and DOA estimation performance.}

We proceed by presenting the observation model in Section~\ref{sec:model} and the proposed PDP approach for DOA estimation in Section~\ref{sec:propose}. Simulation results are presented in Section~\ref{sec:Simu} before drawing the conclusion of the paper in Section~\ref{sec:conc}.

\section{Observation Model}
\label{sec:model}
We consider a complex sinusoidal source signal, with a frequency $f$ and amplitude $A$, $s(t)=Ae^{-j2\pi f t}$ in the \emph{far field}~\cite{12-farfield} of a non-uniform linear array of $N$ antennas. The source impinges on the array from a direction $\theta \in [-\pi/2, \pi/2]$ rad. Let $d_{uv}$ denote the distance between a pair of the array antennas ($u$ and $v$) normalized by $\lambda/2$, where $\lambda$ is the signal wavelength. The received signal (vector) at time $t$ can be modeled as~\cite{doa_model_coprime}
\begin{equation}
\xv(t) = \av(\theta) s(t) + \wv(t),
\label{eq_2}
\end{equation}
where $\av(\theta) = [1, \ \ e^{-j\pi d_{12}\text{sin}(\theta)},\ ..., \ e^{-j\pi d_{1N}\text{sin}(\theta)}]^T$ is the array steering vector, $(\cdot)^T$ indicates the transpose operation, and $\wv(t)$ is a vector of {additive white Gaussian noise (AWGN)}.

The \emph{wrapped phase-difference} (WPD) across an antenna pair, $u$ and $v$, can be estimated from the $u$-th and $v$-th elements of $\xv$ as
\begin{equation}
\hat\psi_{uv}(t) = \mathrm{angle}(x_u(t) \cdot x_v^{*}(t)  ) \in[-\pi, \pi),
\label{phase_estimation}
\end{equation}
where $(\cdot)^*$ is the complex conjugate operation. For simplicity, and without loss of generality, we will focus on single-snapshot scenarios. Hence, we will drop the time variable $t$.

To develop our proposed method, we start from noise-free WPD observations, $\psi_{uv}(\theta)$. These are related to the \emph{actual phase-difference}, $\phi_{uv}(\theta) = \pi d_{uv} \sin(\theta)$ through
\begin{equation}
\psi_{uv}(\theta) = \mathrm{mod}({\phi_{uv}(\theta)}+\pi,2\pi) - \pi = \pi d_{uv} \sin(\theta) - 2\pi q_{uv} ,
\label{eq_1}
\end{equation}
where $\mathrm{mod}(\cdot, \cdot)$ is the modulus operation. The value of $q_{uv}$ can be obtained as
\begin{equation}
q_{uv} = \mathrm{round}\left(\frac{ \pi d_{uv} \sin(\theta) }{2\pi}\right),
\label{wrapped PD}
\end{equation}
where $\mathrm{round(\cdot)}$ is the integer rounding operation.

Based on (3), we observe that estimating the DOA from $\psi_{uv}(\theta)$ requires knowledge of the integer $q_{uv}$, which is not available since we use (2) to estimate $\psi_{uv}(\theta)$. When $d_{uv} \leq 1$, $q_{uv} = 0$ for any $\theta$. However, for $d_{uv} > 1$, the latter result is not guaranteed, except for a specific range of $\theta$ values. Since $\theta$ is unknown, $\psi_{uv}(\theta)$ will always be \emph{ambiguous} for $d_{uv} > 1$, {which is the case for most of the antenna pairs in a NULA.} 

\section{The Proposed PDP Algorithm}
\label{sec:propose}
% \subsection{Proposed DOA estimation algorithm}
\subsection{Wrapped Phase-Difference Pattern (WPDP)}
For an arbitrary source location $\theta \in [-\pi/2, \pi/2]$, using \eqref{eq_1}, we can compute the WPD across receiver pairs to create a WPD vector $\psiv(\theta)=[\psi_{uv}(\theta)]^T, u, v \in\{1,\cdots N\}, u<v$. Assuming that we utilize $M \le {N\choose2}$ antenna pairs, we can simplify the notations and write $\psiv(\theta)=[\psi_m(\theta)]^T$, and $\qv(\theta)=[q_m(\theta)]^T, m=1,\cdots,M$. We can also arrange the inter-antenna distances that correspond to $\psiv(\theta)$ in a vector $\dv=[d_m]^T$.

Now, let us think of $\psiv(\theta)$ as a point in an $M$-dimensional space. From \eqref{eq_1}, and for $\theta=0$, we can see that $\psi_m(\theta)=0, q_m(\theta) = 0, \forall m\in\{1,\cdots,M\}$. {By gradually increasing $\theta$ starting from $\theta=0$, we can see that all $\psi_m(\theta)$ increase linearly with $\sin(\theta)$. The entries of the vector $\qv(\theta)$ remain constant (all zeros) up to a certain $\theta$ value at which the entry corresponding to the largest inter-antenna spacing will have an increment of $+1$. Then, again, $\qv(\theta)$ will remain constant until another entry changes its value. The elements of $\qv(\theta)$ will successively change their value until we reach $\theta=\pi/2$. A similar phenomenon is observed when $\theta$ is varied in the negative direction starting from zero--the entries of $\qv(\theta)$ successively be incremented by $-1$.}        
 This process creates different intervals of $\theta$, each interval with a distinct vector $\qv$ that remains unchanged throughout that interval. Let us denote these intervals as $\Theta_k, k \in\{1,\cdots, K\}$. For any $\theta_a, \theta_b \in \Theta_k$,
 %Depending on the corresponding inter-antenna  spacing $d_m$, an entry of $\qv(\theta)$ may repeatedly increase by a value equal to \emph{one} as $\theta$ increases. The opposite happens when $\theta$ changes in the negative direction.
 \begin{equation}
 \qv(\theta_a) =  \qv(\theta_b) = \qv_k.
\label{q}
\end{equation}
Based on \eqref{eq_1} and \eqref{q}, we can write
\begin{equation}
\delta_{ab} =  \psiv(\theta_a) - \psiv(\theta_b)=\pi \dv[\sin(\theta_a)-\sin(\theta_b)],
\label{intervals}
\end{equation}
% \begin{equation}
% \tv =  \frac{\psiv(\theta_a) - \psiv(\theta_b)}{||\psiv(\theta_a) - \psiv(\theta_b)||_2} =  \frac{\dv}{||\dv||_2}.
% \label{intervals}
% \end{equation}

%The entries of $\qv(\theta)$ remain the same (equals to $\qv_k$) for all $\theta \in \Theta_k$, and so~\eqref{intervals} is valid for any choice of $\{\theta_a, \theta_b\} \in \Theta_k$. 
which indicates that each continuum given by $\psiv(\Theta_k), k \in\{1\cdots, K\}$, is a straight line; and that all the $K$ straight lines point in the same direction $\dv/||\dv||_2$, where $||\cdot||_2$ is the Euclidean norm. That is, we have $K$ parallel lines in $M$-dimensional space, with $K$ given by~\cite{pdp_eusipco2019}
\begin{equation}
\begin{split}
    K = 2\sum^{M}_{i=1}\mathrm{ceil}\left(\frac{\psi_i(\pi/2) - \pi}{2\pi} \right) + 1,
    \label{eq:number_projection_points}
\end{split}
\end{equation}
where $\mathrm{ceil}(\cdot)$ returns the nearest integer greater than or equal to the argument. {These $K$ lines result from \emph{abrupt changes} in the linear relationship between the entries of the vector $\psiv(\theta)$ that occur when an entry of $\qv(\theta)$ changes its value.} We refer a  plot of $\psiv(\theta)$ as a \emph{wrapped phase-difference pattern} (WPDP). 

\begin{figure}[htbp]
\centering
\includegraphics[width=2.07 in]{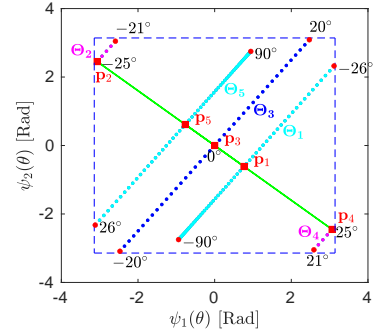}
\caption{An example of a WPDP for $\rv = [0, 2.3, 5.18]$, $M=2$.}
\label{fig-2-wpdp-example}
\end{figure}

An illustration of a WPDP for an array of 3 elements is depicted in Fig.~\ref{fig-2-wpdp-example}. We use $M=2$ and $\psiv(\theta) = [\psi_{12}(\theta) , \psi_{23}(\theta) ]^T$. The inter-antenna  spacing vector is $\rv = [0, 2.3, 5.18]$ (relative to antenna-1). We can see $K\!=\!5$ WPD lines displayed together with the corresponding projection points, $\pv_k, k\!=\!1,\cdots,5$ (will be discussed shortly). {These five lines represent the relationship between the entries of $\psiv(\theta)$ as $\theta$ changes (see \eqref{eq_1}). Sample $\theta$ values (in degrees) are indicated.}       

%We note that for the middle line ($\theta \in[-20^\mathrm{o}, 20^\mathrm{o}]$), $\psiv(\theta) = \phiv(\theta)$, i.e., no phase-difference wrapping is occurring. For all the points located on the other lines, $\psiv(\theta) \neq \phiv(\theta)$ due to phase wrapping. 

%For this line, $\psi_2$  reaches the value $\pi$ at around $\theta = 20^\mathrm{o}$, resulting in a jump or \emph{wrapping} when $\theta$ increases further. 
%The direction of the WPD lines $\dv/||\dv||_2$ can be obtained with $\dv = \psiv(20^\circ)-\psiv(0^\circ)$.  

% Some compensation is needed to be able to perform DOA estimation. The following subsection addresses this issue.

Using simple geometry, we can see that all the WPD lines $\psiv(\Theta_k)$ (dotted lines) are perpendicular to a hyperplane (solid green line in this 2-dimensional WPDP) that contains the origin. This hyperplane is formed by the points $\psiv$ that satisfy
\begin{equation}
\dv^T \psiv = d_1 \psi_1  + d_2 \psi_2  + \cdots + d_M \psi_M = 0.
\label{hyperplane}
\end{equation}
Each line $\psiv(\Theta_k)$ has a \emph{projection point}, $\pv_k$, at which the line intersects with the hyperplane. In the following discussion, we show how to compute these projection points.

Given a noise-free WPD vector $\psiv(\theta)$, the distance between this vector, treated as a point in $M$-dimensional space, and the hyperplane \eqref{hyperplane}, is given by
\begin{equation}
\mathrm{dist}\left( \psiv(\theta), \dv^T\psiv=0 \right) = \frac{\dv^T\psiv(\theta)}{||\dv||_2}.
\end{equation}
The projection point $\pv(\theta)$ of $\psiv(\theta)$ on the hyperplane along the direction $\dv/||\dv||$ can be obtained as
\begin{equation}
\pv(\theta) = \mathrm{project}(\psiv(\theta)) = \psiv(\theta) - \frac{\dv^T \psiv(\theta)}{||\dv||_2}\cdot\frac{\dv}{||\dv||_2}.
\label{eq-5projecting}
\end{equation}
It is obvious that all points on the same WPD line are projected on the same point in the projection hyperplane. However, for an observed noisy wrapped phase-difference vector $\hat \psiv$, (\ref{eq-5projecting}) returns a perturbed projection point, possibly $\hat\pv \notin \{\pv_k$\}. In this case, we pick the nearest projection point $\pv_z$, where
\begin{equation}
    z = \operatorname*{arg\,min}_{k}||\pv_k-\hat\pv||_2.
    \label{eq:closest_p}
\end{equation}
Now, the unbiased WPD $\tilde \psiv$, which is the nearest point on the line with the projection point $\pv_k$ can be estimated as
\begin{equation}
   \tilde\psiv =\pv_z + \frac{\dv^T \hat\psiv}{||\dv||_2}\cdot\frac{\dv}{||\dv||_2}.
   \label{eq:unbiased_wpd}
\end{equation}

Based on \eqref{q}, all points on the $k$-th wrapped phase-difference (WPD) line can be compensated/unwrapped with the same \emph{unwrapping vector} $\hv_k = 2\pi \qv_k$, i.e., $\phiv(\theta) = \psiv(\theta) +  \hv_k, \forall \theta \in \Theta_k
$. Hence, the estimated \emph{unwrapped} phase difference can be obtained as
\begin{equation}
\hat\phiv = \hat\psiv +  \hv_z.
\label{unwrapping}
\end{equation}
A procedure to compute the unwrapping vector $\hv_k$ and the projection points $\pv_k$ will be detailed in the next subsection.

% A new line is started at a point vertically aligned with the endpoint of the middle line (near the bottom-right corner with a mark $21^\mathrm{o}$  at the start). 
% For this line, and the rest of the points located on other lines, $\psiv(\theta) \neq \phiv(\theta)$ due to phase wrapping. Some compensation is needed to be able to perform DOA estimation. The following subsection addresses this issue.

\subsection{Computing the Unwrapping Vectors and Projection Points}
The unwrapping vector $\hv_k$ can be obtained by tracing the WPD lines. Together with their projection points, these lines are easily identified by their (known) direction unit vectors and starting points. We can start from the point $\psiv(-\pi/2)$, which, let us say, falls on the first line. The point where this line intersects with the M-cube whose boundaries are $-\pi$ and $\pi$ can easily be calculated. The intersection point determines the next line's starting point, which is obtained by wrapping the coordinate of $\psiv$ that crosses the cube's surface. 
% This way, we can obtain all the projection points $\pv_k$ and their corresponding compensation vectors $\hv_k$.
A pseudocode for calculating $\pv_k$ and $\hv_k$ is listed in Algorithm~\ref{alg_initialization}, where $\phiv_0$, $\psiv_0$ and $\hv_0$ are intermediate variables. {The procedure in Algorithm~\ref{alg_initialization} is performed completely offline, which reduces the online complexity of the proposed algorithm.}

\begin{algorithm}[!htb]
\caption{-- Calculate $\hv_k$, $\pv_k$}
\label{alg_initialization}
\begin{algorithmic}[1]
% \State $j \leftarrow 1$
\State $\phiv_{0} \leftarrow \pi \dv \sin(-\pi/2)$, $\phiv_{max} \leftarrow \pi \dv \sin(\pi/2)$
% \State $\phiv_{max} \leftarrow \pi \dv \sin(\pi/2)$
\State $j \leftarrow 1$, $ \hv_{0} \leftarrow \phiv_{0} - \mathrm{wrap}(\phiv_{0})$
\While{$\phiv_{0}(m)<\phiv_{max}(m)$ for all $m = [1,2,...,M]$}
%   \State $\psiv_{b,j} \leftarrow \psiv_{0}$
  \State $\hv_j \leftarrow \hv_{0}$
  \State $\pv_j \leftarrow \mathrm{project}(\psiv_{0})$
  \State $j \leftarrow j+1$
  \State $i \leftarrow \operatorname*{arg\,min}_{i} [(\pi - \psiv_{{c}}(i))/d_i]$
  \State $\psiv_{0} \leftarrow \dv(\pi-\psi_{0}(i))/d_{i} + \psiv_{0}$
  \State $\psiv_{0}(i) \leftarrow \psiv_{0}(i)-2\pi$
  \State $\hv_{0}(i) \leftarrow \hv_{0}(i) + 2\pi$
  \State $\phiv_{0} = \psiv_{0} + \hv_{0}$
\EndWhile
\State
\Return $\hv_k, \pv_k (k\in\{1,2,...,K = j-1\})$
\end{algorithmic}
\end{algorithm}

\subsection{PDP DOA Estimation Algorithm}
% The proposed DOA algorithm starts by initializing $\hv_k$ and $\pv_k$ using algorithm~\ref{alg_initialization}. 
% Then, for 
Given a noisy WPD $\hat\psiv$, the estimated projection point $\hat\pv$ can be computed using~\eqref{eq-5projecting} and~\eqref{eq:closest_p}. Then, the unbiased WPD $\tilde \psiv$ can be obtained using~\eqref{eq:unbiased_wpd}. Next, the estimated unwrapped phase-difference vector $\hat \phiv$ can be obtained using~\eqref{unwrapping}. Finally, the DOA of the source can be calculated using~\eqref{eq_1}. A pseudocode of the proposed algorithm is listed in Algorithm~\ref{alg_pdp_pseudocode} (Matlab codes available in~\href{https://github.com/chenhui07c8/DOA-AOA-algorithms/tree/master/4\%20PDP\%20for\%20NULA}{\textit{\textbf{Github}}}).

%An example of implementing the PDP algorithm is shown in Fig.~\ref{fig-2-wpdp-example}, where the target DOA is at $22^\circ$. $\hat\psiv(22^\circ)$, $\tilde \psiv(22^\circ)$, $\hat \phiv(22^\circ)$ are the estimated , the unbiased WPD and the estimated unwrapped phase-difference vector, respectively. This is just an illustration of DOA estimation for two antennas; however, the algorithm applies to more pairs of antennas.

\emph{Remark~1:}
In Algorithm~\ref{alg_pdp_pseudocode}, we note that the bulk of the computational complexity lies in Step-1. This step needs to be performed only once at the initial setup (offline). The rest of the algorithm's (online) steps involve simple computations. This, along with its grid-less nature, greatly enhances the online computational complexity of the proposed algorithm.

\begin{algorithm}[!htb]
\caption{-- PDP DOA Estimation Algorithm}
\label{alg_pdp_pseudocode}
\begin{algorithmic}[1]
\State Initialize $\hv_1$-$\hv_K$, $\pv_1$-$\pv_K$ using Algorithm 1
\State $\hat\psiv \leftarrow \mathrm{angle}(x_u(t_0)\cdot x_v^{*}(t_0))$ for selected pairs  \text{     ~\eqref{phase_estimation}}
\State $\hat\pv \leftarrow \hat\psiv - \frac{\dv^T \hat\psiv}{||\dv||}\cdot\frac{\dv}{||\dv||}$ \text{     ~\eqref{eq-5projecting}}
\State $z \leftarrow \operatorname*{arg\,min}_{k}||\pv_k-\hat\pv||$\text{     ~\eqref{eq:closest_p}}
\State $\tilde\psiv \leftarrow \pv_z + \frac{\dv^T \hat\psiv}{||\dv||}\cdot\frac{\dv}{||\dv||}$\text{     ~\eqref{eq:unbiased_wpd}}
\State $\hat\phiv \leftarrow \tilde\psiv + \hv_z$ \text{     ~\eqref{unwrapping}}
\State $\hat\theta \leftarrow \mathrm{sin}^{-1}\left(\frac{\hat\phiv_1}{\pi d_1}\right)$\text{     ~\eqref{eq_1}}
\State
\Return $\hat\theta$
\end{algorithmic}
\end{algorithm}

\begin{figure}[htb]
\begin{minipage}[b]{0.48\linewidth}
  \centering
  \centerline{\includegraphics[width=4.4cm]{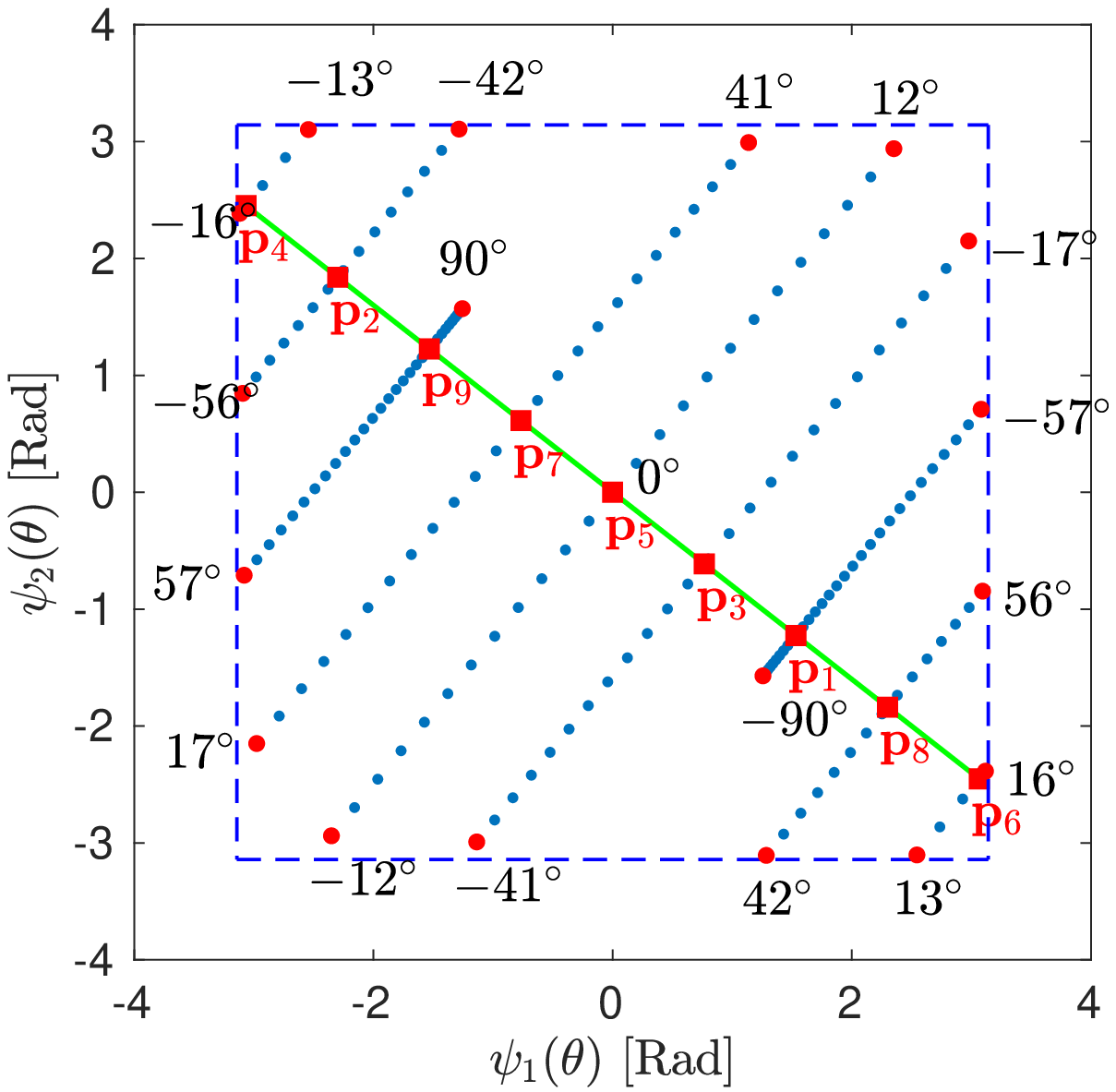}}
%  \vspace{1.5cm}
  \centerline{(a) $\rv = [3.6, 8.1]$} \medskip
\end{minipage}
\hfill
\begin{minipage}[b]{0.48\linewidth}
  \centering
  \centerline{\includegraphics[width=4.4cm]{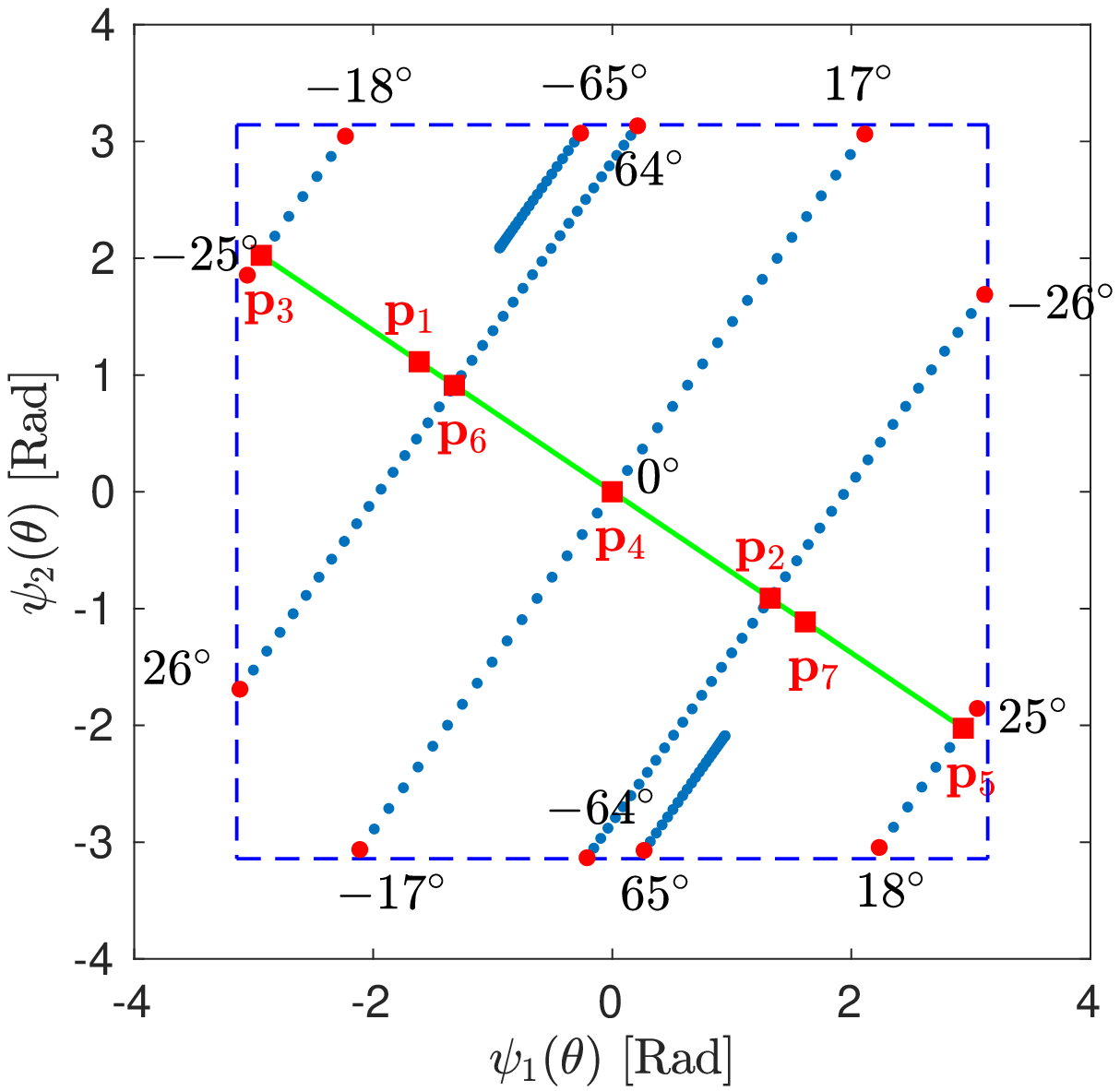}}
%  \vspace{1.5cm}
\centerline{(b) $\rv = [2.3, 5.6]$}\medskip
\end{minipage}
\caption{Visualization of two different WPDPs.}
\label{fig-3}
\end{figure}

\emph{Remark~2:}
Noise perturbation in WPD tends to drift a noisy WPD point away from its original (noise-free) WPD line. The proposed PDP algorithm associates a noisy point with the closest projection point. Depending on the noise level, and the distance between the projection points, this might lead to a wrong (hard) decision and erroneous DOA estimation. 
% It is obvious that the impact of noise depends on the distances between the projection points; the farther apart the projection points, the less severe the impact of noise in the determination of the projection point. 
For a specific signal frequency, the distance between the projection points is solely determined by the array layout, which can easily be deduced from \eqref{eq-5projecting}. 
%The characteristics of different array configurations based on their WPDP structures can provide an interesting tool for analyzing these arrays' behavior for different source locations. 
% The exploration of these characteristics requires abundant space and is left out of the scope of this paper. To conclude this remark, 
We illustrate the impact of array configuration in Fig.~\ref{fig-3}, which shows WPDP examples for two different array configurations with the same number of elements.
From the figure, we can see a significant difference in WPDP structures for the two arrays, as reflected in the number of projection points and the inter-point distances.
% \textcolor{blue}
As an example, it is expected that recovering a projection point correctly from noisy observations to be easier for $\pv_4$ in Fig.~\ref{fig-3}~(b) compared to the rest of the projection points in both Fig.~\ref{fig-3}~(a) and (b).
On the other hand, in situations where two WPD lines overlap, the proposed algorithm, or any other DOA estimation algorithm, will fail to identify the source location. This happens when the array configuration is \emph{ambiguous}~\cite{Manikas1998}.

\begin{figure*}[htb]
\begin{minipage}[b]{0.3\linewidth}
  \centering
  \centerline{\includegraphics[width=5.6cm]{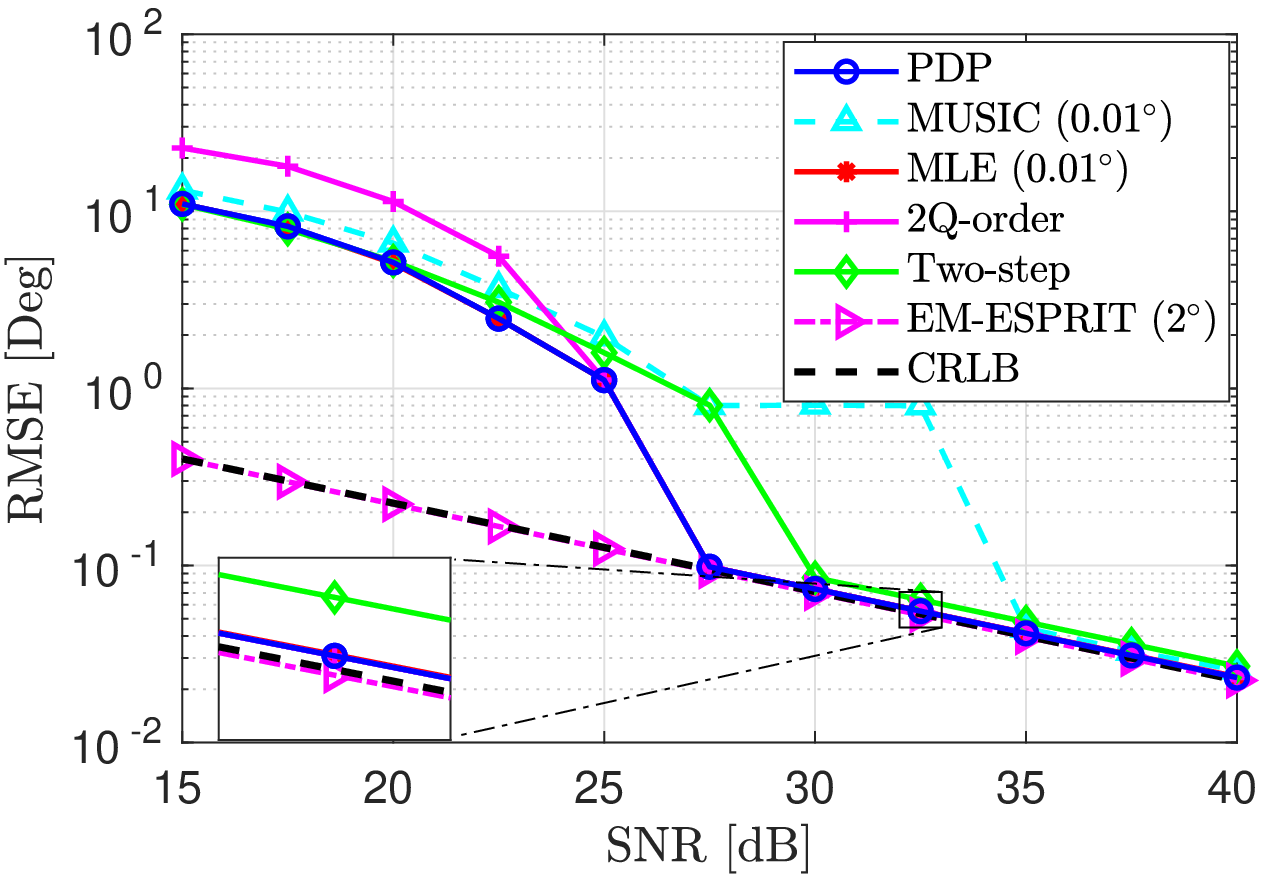}}
%  \vspace{1.5cm}
  \centerline{(a) $\rv_{1\text{-}3}$ ($K=21$)} \medskip
\end{minipage}
\hfill
\begin{minipage}[b]{0.3\linewidth}
  \centering
  \centerline{\includegraphics[width=5.6cm]{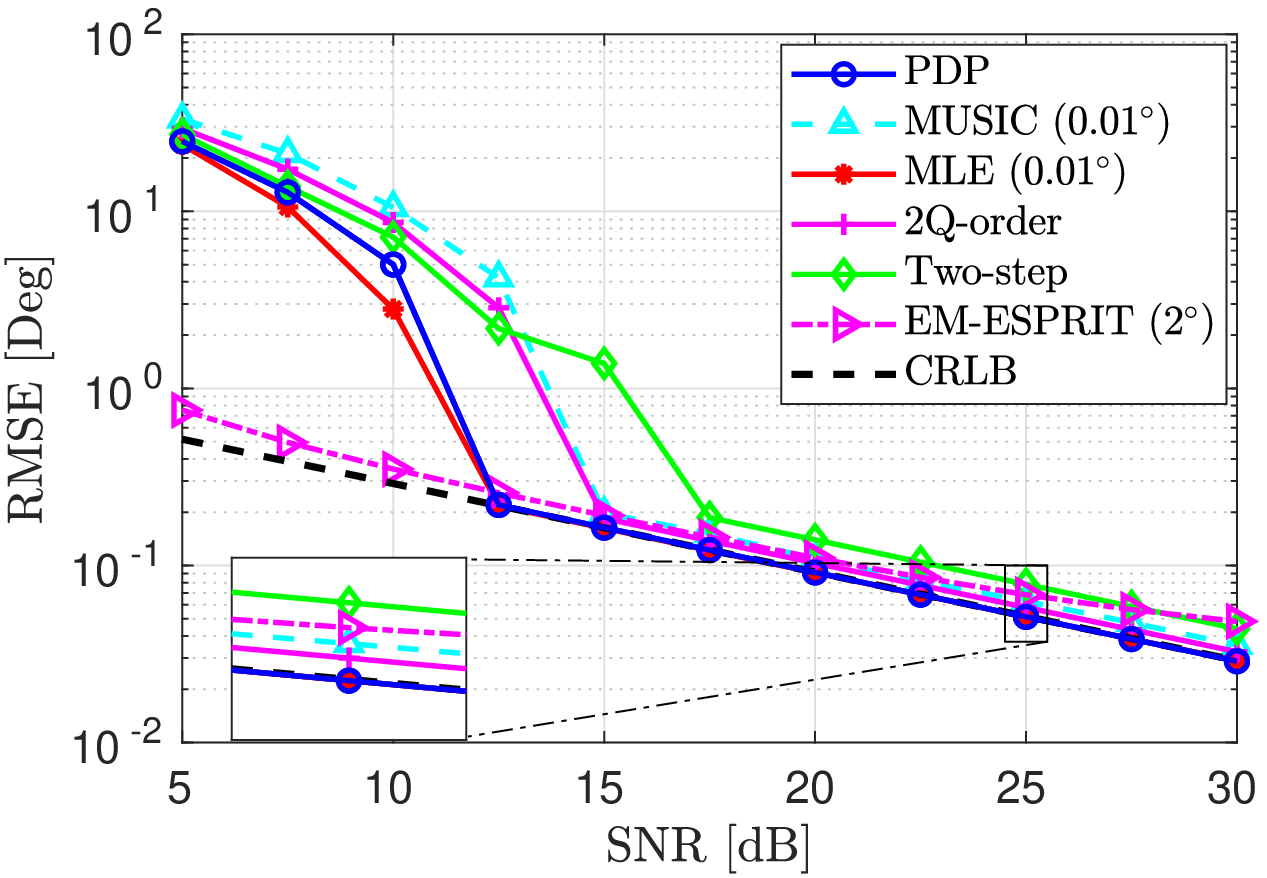}}
%  \vspace{1.5cm}
\centerline{(b) $\rv_{1\text{-}5}$  ($K=109$)}\medskip
\end{minipage}
\hfill
\begin{minipage}[b]{0.3\linewidth}
  \centering
  \centerline{\includegraphics[width=5.6cm]{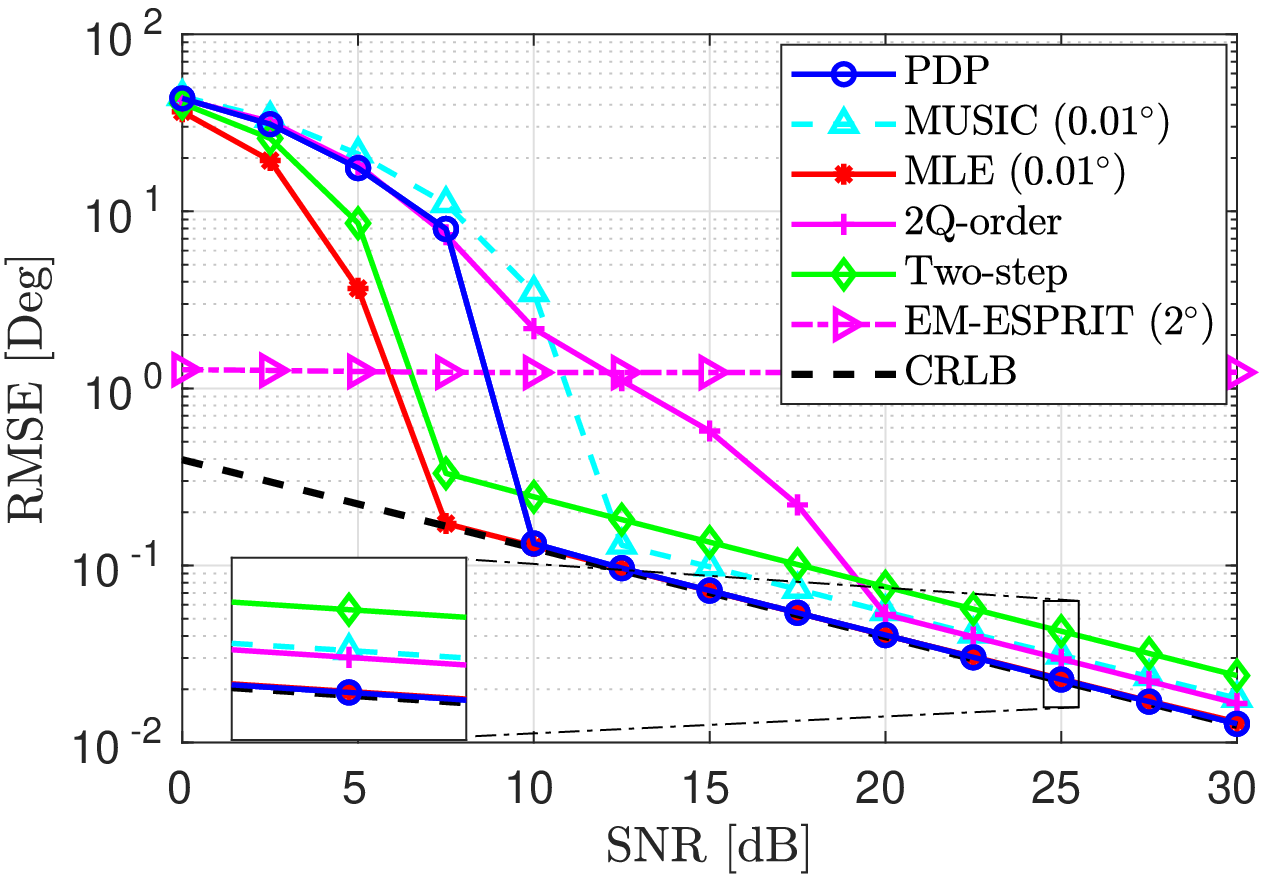}}
%  \vspace{1.5cm}
\centerline{(c) $\rv_{1\text{-}8}$ ($K=515$)}\medskip
\end{minipage}
% S = [0 10 21 33]*lambda/2*0.5;
\begin{minipage}[b]{0.3\linewidth}
  \centering
  \centerline{\includegraphics[width=5.6cm]{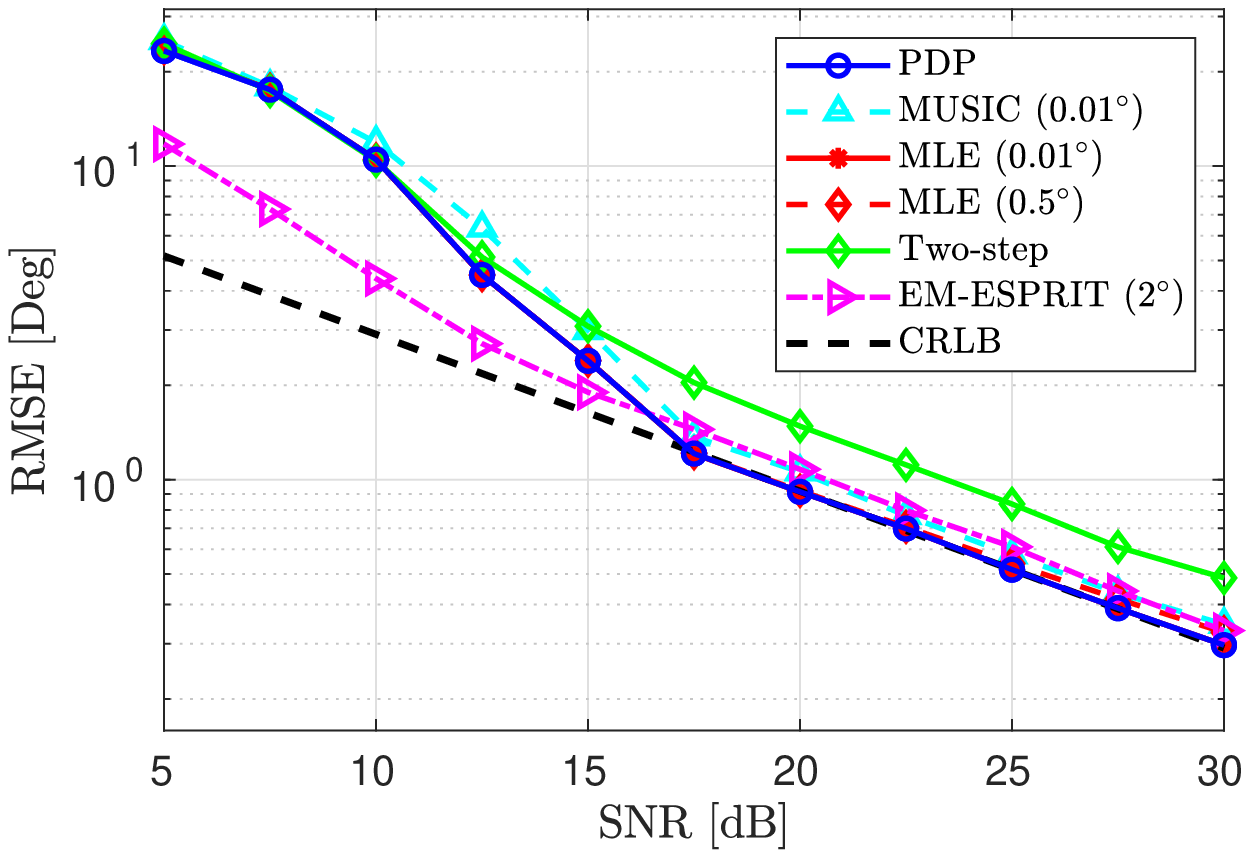}}
%  \vspace{1.5cm}
\centerline{(d) $\rv_{2\text{-}3}$ ($K=5$)}\medskip
\end{minipage}
\hfill
\begin{minipage}[b]{0.3\linewidth}
  \centering
  \centerline{\includegraphics[width=5.6cm]{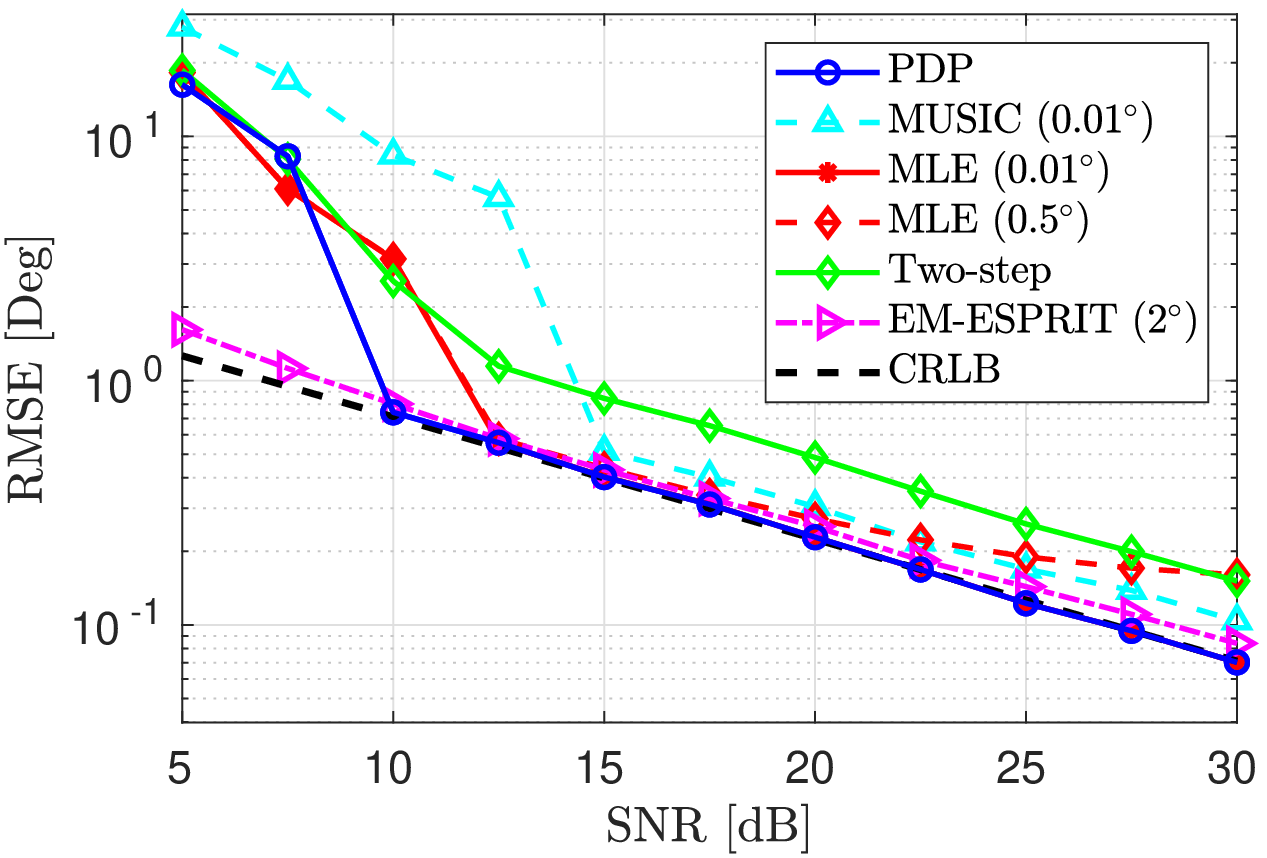}}
%  \vspace{1.5cm}
\centerline{(e) $\rv_{2\text{-}5}$  ($K=41$)}\medskip
\end{minipage}
\hfill
\begin{minipage}[b]{0.3\linewidth}
  \centering
  \centerline{\includegraphics[width=5.6cm]{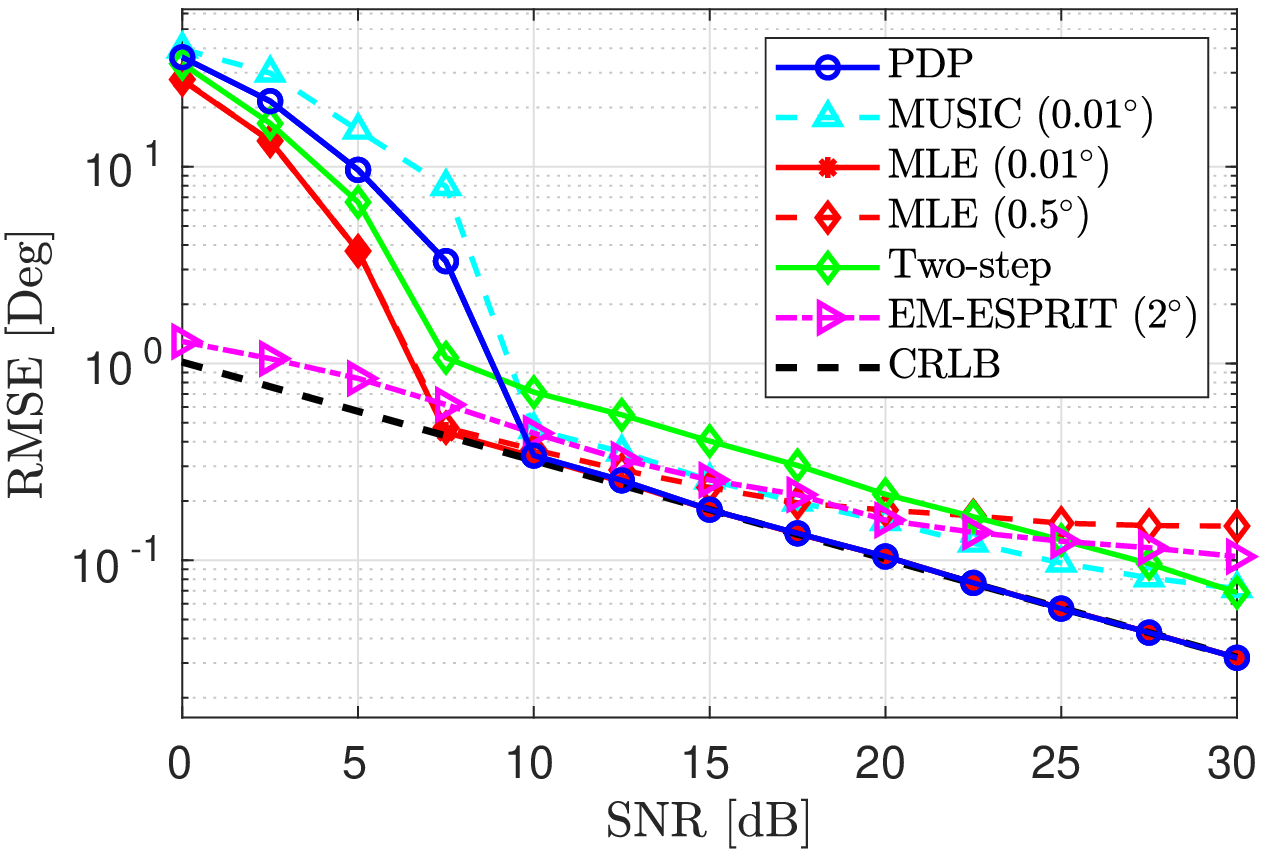}}
%  \vspace{1.5cm}
\centerline{(f) $\rv_{2\text{-}8}$  ($K=201$)}\medskip
\end{minipage}
\caption{RMSE versus SNR for different algorithms tested over six different array configurations. $K$ is the number of projection points in each setup.}
\label{simulations}
\end{figure*}

% \subsection{MSE Approximation Simulation Results}
% In MSE approximation simulation, the parameters are chosen the same as Fig.~\ref{fig-2-wpdp-example}. antenna pairs $\psiv =[\psi_{12}, \psi_{23}, \psi_{13}]$ are chosen for the PDP algorithm.

% The simulation of MSE estimation for different DOAs and SNRs is shown in Fig.~\ref{SIM-1}. The estimation of MSE using~\eqref{eq_mse_estimation} fits the simulation result well when SNR is high. For low SNR, the noisy phase estimation error might be greater than $L_{m}$, and hence, the approximation is invalid. However, this is a single-shot phase estimation, and a better estimation can be obtained by utilizing multiple snapshots.
% \begin{figure}[!htb]
% \centering
% \includegraphics[width=2.2 in]{fig-5-1.pdf}
% \caption{MSE estimation for different DOA and SNR. (N=2)}
% \label{SIM-1}
% \end{figure}

\section{Performance Evaluation}
\label{sec:Simu}
In this section, we evaluate the performance of the proposed PDP algorithm given in Algorithm~\ref{alg_pdp_pseudocode} along with several benchmark methods. The benchmark methods are the 2Q-order algorithm~\cite{2q-order}, {the two-step offset correction method~\cite{ma2019off}}, 
{the expectation maximization ESPRIT (EM-ESPRIT) algorithm for NULAs~\cite{el2007esprit},} 
the MUSIC algorithm~\cite{13-MUSIC}, and the MLE~\cite{crlb_threshold_region}. We also compare with Cram\'er-Rao lower bound (CRLB) given by~\cite{crlb_threshold_region}
\begin{equation}
\text{CRLB} = \frac{1}{2\pi^2NSU\sin(\theta)},\ \  U = \frac{1}{N}\sum^{N}_{n=1}\left(r_n - \bar{\rv}\right)^2,
\end{equation}
where $\theta$ is the source location, $S$ is the linear signal-to-noise ratio (SNR), and $\rv$ is an $N\times 1$ antenna spacing vector, $r_n$ and $\bar{\rv}$ are the $n$-th element and the average value of $\rv$, respectively.

\begin{table}[htbp]
\caption{Online Computational Complexity for Different Algorithms}
\label{table:complexity}
\centering
\begin{tabular}{c |c | c c}
\hline
\textbf{Algorithms}  & \textbf{Off-Grid}  & \textbf{Number of Multiplications}  \\
\hline
{PDP}             & Yes   &  $(K\!+\!6)\frac{N(N-1)}{2}$ \\
\hline
{Two-step}        & Yes     &  $(K_c\!+\!1)N$\\
\hline
{2Q-order}        & Yes    &  $N$\\
\hline
{EM-ESPRIT}     & Yes & $K_i(2N_v^2 + \frac{16}{5}N_v^3)$ \\
\hline
{MUSIC}     & No & $\frac{16}{5}N^3 + (K_c\!+\!K_f\!+\!\frac{N}{2})N(N\!+\!1)$ \\
\hline
{MLE}       & No & $(K_c\!+\!K_f\!+\!\frac{1}{2})N(N\!+\!1)$\\
\hline
\end{tabular}
\end{table}

\begin{table}[htbp]
\caption{Online Multiplications required with Different Configurations}
\label{table:multiplications_needed}
\centering
\begin{tabular}{c |c | c |c | c |c |c}
\hline
\textbf{Algorithms}  & $\rv_{1\text{-}3}$  &  $\rv_{1\text{-}5}$ &  $\rv_{1\text{-}8}$ &  $\rv_{2\text{-}3}$ &  $\rv_{2\text{-}5}$ &  $\rv_{2\text{-}8}$  \\
\hline
{PDP}   & $81$ & $1150$ & $14588$ & $33$ & $470$ & $5796$  \\
\hline
{MLE}   & $8670$ & $21675$ & $52020$ & $8670$ & $21675$ & $52020$\\
\hline
\end{tabular}
\end{table}
\subsection{Computational Complexity Analysis}
{We use the number of multiplication operations to characterize each algorithm's computational complexity, as summarized in Table~\ref{table:complexity}. The symbol $K$ denotes the number of projection points of the PDP algorithm, which is array-layout dependent, as given in (7). {The symbols $N_v$ and $K_i$ are the virtual array size and the number of iterations for EM-ESPRIT.} For grid-search-based methods, the parameters $K_c$ and $K_f$ are the numbers of grid points used in the coarse search and fine search, respectively. From the table, we observe that 2Q-order has the lowest computational cost of all algorithms, which depends only on $N$. However, this algorithm requires specific array configurations. The two-step method has the second-lowest complexity, which is $\propto K_c N$. For PDP, {EM-ESPRIT}, MUSIC, and MLE, the complexity comparison depends on the values of $K$, {$K_i$}, $K_c$, and $K_f$. PDP can outperform the former two algorithms in computational complexity when $K$ is small compared to {$K_i$ and} $K_c\!+\!K_f$.}
%The complexity of PDP can be reduced by judiciously choosing a sufficient subset of all the antenna pair combinations to achieve adequate performance.}

\subsection{Simulation Results}
{In our simulations, we utilize two array configurations, namely, $\rv_1\!=\![0, 5, 10.5, 16.5, 23, 30, 37.5, 45.5]$ (setup from~\cite{2q-order}) and $\rv_2\!=\![0,0.4,2.4,4,9.2, 10.4,13.6,16.4]$ (a non-redundant array from~\cite{vertatschitsch1986nonredundant}). From these arrays, we create six scenarios using three subarrays of each of $\rv_1$ and $\rv_2$ with different number of antennas $N\!=\!3,5,8$ (e.g., $\rv_{1\text{-}3} = [0, 5, 10.5]$ is a subarray of the first three elements of $\rv_1$).} For each scenario, we plot the \emph{root mean squared error} (RMSE) versus SNR calculated from 1000 simulation trials at each SNR value. In each trial, the source location is generated randomly from a uniform distribution between $39.5^\circ$ to $40.5^\circ$. For the proposed method, $M = {N\choose 2}$ phase-difference estimates are computed using~\eqref{phase_estimation}. 
For the MUSIC and MLE algorithms, the search is implemented in two stages~\cite{crlb_threshold_region}--a coarse search in the interval $[-70^\circ, 70^\circ]$ using a $0.2^\circ$ step followed by a fine search using a $0.01^\circ$ step.
{An initial estimation is needed for the EM-ESPRIT~\cite{el2007esprit}, which is chosen randomly from $[\theta\!-\!2^\circ, \theta\!+\!2^\circ]$. Also, we set $K_i=20$.}
In all simulation trials, and for all methods, a single snapshot is used to estimate the source location. The RMSE performance for all scenarios is presented in~Fig.~\ref{simulations}. {The 2Q-order algorithm only works for $\rv_1$ due to a specific geometry requirement, and hence is not shown in the results based on $\rv_2$ (Fig.~\ref{simulations} (d)-(f)).} 

{For the scenarios of Fig.~\ref{simulations}, the proposed PDP algorithm mostly matches the RMSE of the MLE (fine search resolution of $0.01^\circ$), with some deviations that occur at low SNRs, especially for relatively large arrays. Both PDP and MLE ($0.01^\circ$ resolution) achieve the CRLB at high SNRs in all the tested scenarios. The other benchmark methods tend to lack \emph{consistency} as they deviate from the CRLB in the high SNR regime (exceptions are the 2Q-order and EM-ESPRIT algorithms in Fig.~\ref{simulations}~(a)). The EM-ESPRIT algorithm tends to outperform the rest of the methods at low SNRs. However, this can be attributed to the extra information available to this algorithm in the form of a good initial point. The performance of the on-grid methods (MUSIC and MLE) highly depends on the search step. As an example, the MLE performance using only a (coarse) search step of $0.5^\circ$ deviates from the CRLB in scenarios (e) and (f).}      

%For the rest of the configurations, the proposed algorithm does not match the MLE and two-step methods in some low SNR scenarios but is still close to the CRLB in the high SNR region.
%The EM-ESPRIT algorithm is better than the other benchmarks with a good initial estimation in low SNR region. However, it does not work well with the increase of SNR and array size.} The performance of the on-grid methods (MUSIC and MLE) highly depends on the search step. As an example, the MLE performance using only a (coarse) search step of $0.5^\circ$ deviates from the CRLB in the scenarios (e) and (f). 

{As for computational complexity, the two-step and 2Q-order methods require fewer computations than all the other methods. Based on the above discussion, these methods, together with MUSIC and EM-ESPRIT, offer inferior performance on average (compared to PDP and MLE).} Therefore, in Table~\ref{table:multiplications_needed}, we show the actual number of multiplications for only PDP and MLE (with $0.01^\circ$ search step) in all six scenarios. From the table, we observe a substantial advantage for the proposed PDP algorithm. This advantage is attributed mainly to the avoidance of grid search and that the proposed algorithm performs a good amount of its computations offline.

\section{Conclusion}
\label{sec:conc}
A phase-difference projection (PDP) direction of arrival (DOA) algorithm is proposed. The proposed algorithm projects the phase-difference observations measured across antenna pairs on a predefined hyperplane determined by the array geometry. Based on this projection, DOA estimation can be achieved in a simple and computationally efficient manner. Simulation results show that the proposed algorithm can match maximum likelihood estimation while maintaining a significant computational-complexity advantage.

\ifCLASSOPTIONcaptionsoff
\newpage
\fi

% \end{thebibliography}
\bibliographystyle{IEEEtran}
\bibliography{chen_WCL2021-0665}

\end{document}

%% file: macros.tex
\long\def\comment#1{}

\newfont{\bbb}{msbm10 scaled 700}

\newfont{\bb}{msbm10 scaled 1100}

\newcommand{\av}{{\bf a}}

\newcommand{\dv}{{\bf d}}

\newcommand{\hv}{{\bf h}}

\newcommand{\pv}{{\bf p}}
\newcommand{\qv}{{\bf q}}
\newcommand{\rv}{{\bf r}}

\newcommand{\wv}{{\bf w}}

\newcommand{\xv}{{\bf x}}

\newcommand{\phiv}{\hbox{\boldmath$\phi$}}
\newcommand{\psiv}{\hbox{\boldmath$\psi$}}